\documentclass[letter]{aa} % for the letters
\usepackage{graphicx}
\usepackage{txfonts}
\usepackage{amsmath}
\usepackage{upgreek}
\usepackage{diagbox}
\usepackage[colorlinks=true,linkcolor=blue, citecolor=blue, filecolor=blue, urlcolor=blue]{hyperref}

\begin{document}

   \title{Resonance coupling in spiral arms}

   \subtitle{Patterns for flat rotation curve}

   \author{Alexander A. Marchuk
          \inst{1,2}}

   \institute{Central (Pulkovo) Astronomical Observatory, Russian Academy of Sciences, Pulkovskoye chaussee 65/1, St. Petersburg 196140, Russia
         \and
             Saint Petersburg State University, Universitetskĳ pr. 28, St. Petersburg 198504, Russia\\
             \email{a.a.marchuk+astro@gmail.com}
             }

   \date{Received September 15, 1996; accepted March 16, 1997}

% \abstract{}{}{}{}{}
% 5 {} token are mandatory
 
  \abstract
   {To address questions about the physical nature and origin of spiral arms in galaxies, it is necessary to measure their dynamical properties, such as the angular speed, $\Omega_\mathrm{p}$, or the corotation radius. Observations suggest that galaxies may contain several independent spiral patterns simultaneously. It was shown that so-called non-linear resonance coupling plays an important role in such systems.}
   {We aim to identify cases of independent spiral patterns for galaxies with a flat rotation curve and to investigate what relative pattern velocities, $\mathrm{\Omega^{out}_{p}/\Omega^{in}_{p}}$, they might have for all possible cases of coupling between the main resonances.}
   {We solved equations for the main resonance positions (1:1, 2:1, 4:1) and estimated the ratio $\upvarpi$ of the corotation radii for two subsequent patterns. For six close galaxies with flat rotation curves, we collected the measurements of the corotation radii in the literature, using at least three different methods in each case for credibility. We found at least two independent spiral patterns for each galaxy and measured the $\upvarpi$ ratios.}
   {We found $\upvarpi$ ratios for all possible cases for the main resonances. For three cases, we obtained $\upvarpi>3$, indicating that it would be difficult to fit two or even more spiral patterns in the disc. These ratios have been used to derive the wind-up time for spirals, estimated to be several galactic rotations. We find that three pairs of coupling cases, including those that have been vastly acknowledged in galaxies, namely, $\mathrm{OLR_{in}=CR_{out}}\; \textit{\&} \;\mathrm{CR_{in}=IUHR_{out}}$, have very close $\upvarpi$ ratios; hence, they ought to be found simultaneously, as observed. We find a strongly confirmed apparent resonance coupling for six galaxies and we show that the observed $\upvarpi$ is in agreement with theory. In two of them, we identified a previously unreported form of simultaneous coupling, namely, $\mathrm{OLR_{in}=OUHR_{out}}\; \textit{\&} \; \mathrm{OUHR_{in}=CR_{out}}$. This result was also predicted from the proximity of $\upvarpi$. }
   {}

   \keywords{galaxies: spiral --
                galaxies: fundamental parameters --
                galaxies: structure
               }

   \maketitle
%
%________________________________________________________________

\section{Introduction}

Spiral arms are present in most galaxies \citep{2006MNRAS.373.1389C,2013MNRAS.435.2835W} and as large-scale prominent structures across the disc, they play an essential role in their evolution \citep{2014PASA...31...35D,2016ARA&A..54..667S,2022ARA&A..60...73S}. Such processes as angular momentum transfer or disc heating are crucial for understanding galaxies, thus highlighting the importance of researching spiral arms. However, the key question of their nature -- namely, whether they rotate at an angular speed, $\Omega_\mathrm{p}$, that is similar to that of the disc $\Omega(r)$ --  has not yet been resolved for observations of real objects en masse. Although, in some papers, spiral arms have been linked to the presence of bars or companion satellites \citep{1979ApJ...233..539K,2011MNRAS.414.2498S}, in most  cases (especially in numerical models), spirals are expected to be long-lived density waves with $\Omega_\mathrm{p} \approx \mathrm{const}$ \citep{1964ApJ...140..646L,1989ApJ...338...78B} or rapidly evolving transient features with $\Omega_\mathrm{p} \approx \Omega$, often called dynamic arms \citep{1984ApJ...282...61S,2019MNRAS.489..116S}.
\par
Among many other reasons, this question remains unanswered because spiral arms are probably  not that simple in structure, as originally formulated in the aforementioned theories. It has been shown for both observed galaxies \citep{2009ApJS..182..559B,2014ApJS..210....2F,2009ApJ...702..277M} and $N$-body numerical models  \citep{1999A&A...348..737R,2011MNRAS.417..762Q} that they can contain multiple patterns, rotated with individual $\Omega_\mathrm{p}$. It was also shown theoretically that in such cases individual patterns do not have arbitrary angular speeds; rather, they form a so-called "resonance coupling." Using well-established mathematics for the description of density waves, \citet{1987ApJ...318L..43T} and \citet{1988MNRAS.232..733S} suggested that global modes in stellar discs could be coupled through non-linear interactions, namely, the first wave excites the second one through second-order coupling terms, which are large when resonances coincidence by radii. In their scenario, the corotation radius of an inner pattern overlaps with the inner Lindblad resonance of an outer one. This overlap makes the interaction between the two patterns much more efficient, enabling the transfer of energy and angular momentum between the bar, spiral density wave, and so-called beat waves, which result from the conservation law. This behaviour is in fact a general property of non-linear wave coupling and has also been observed in other fields, such as plasma physics \citep{1982PlPh...24..753T}. \citet{masset1997} theoretically and numerically justified such couplings, not only between bar and spiral, but also between spiral modes. This result has been firmly confirmed via the $N$-body simulations of \citet{1999A&A...348..737R}, which report remarkable coincidences for other resonances, along with  more recent simulations of galactic discs \citep{minchev2012}. 
\par
Despite the fact that resonance coupling has been predicted in theory, demonstrated in models, and formally found in observations of real galaxies, it remains poorly understood. In particular, we do not know which resonances can be coupled, in principle, and which limitations coupling imposes on the velocities of the inner $\Omega_\mathrm{p}^\mathrm{in}$ and outer $\Omega_\mathrm{p}^\mathrm{out}$ patterns. In this Letter we clarify these issues for the simple case of a flat rotation curve and for all possible cases of main resonances, regardless of whether they have been previously demonstrated to be the case in real galaxies or not.

\begin{figure}\includegraphics[width=0.95\columnwidth]{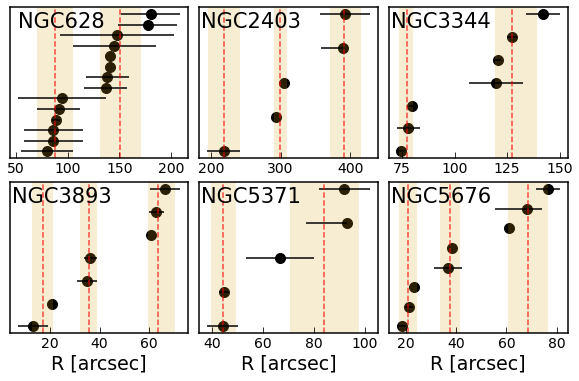}
        \caption{Positions of CR from Table~\ref{tab:crref} (points). Colored vertical areas show boundaries of assumed CR for each pattern, dashed red line shows average value for points within each CR. Vertical axis is shown for illustrative purposes only.}
    \label{fig:crs}
\end{figure}

\begin{figure*}\includegraphics[width=1.95\columnwidth]{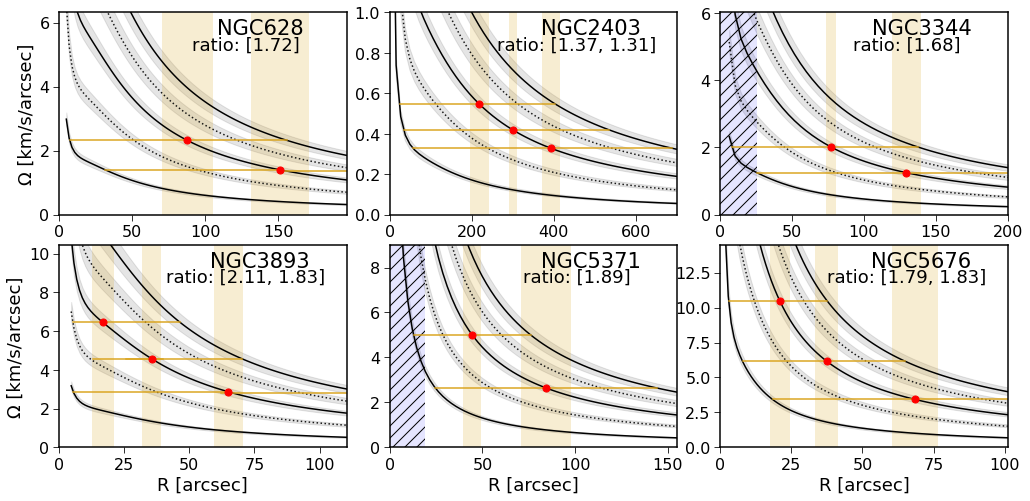}
    \caption{Angular velocity and resonances curves: solid lines show $\Omega$, ILR, and OLR; dotted lines show IUHR and OUHR. Areas filled with gray colour represent error associated with RC. Red points show the average position of each CR. Horizontal lines demonstrate extension of each spiral pattern from it's inner resonance (ILR or IUHR) to the outer one (OLR). Numbers show $\upvarpi$ for consecutive patterns. Blue hatched areas mark bar size from \citet{2015A&A...582A..86H}.}
    \label{fig:omegas}
\end{figure*}

\section{Resonance coupling}

\subsection{Which resonances can be coupled}
\label{sec:first}

We use the standard notations $\Omega$ for angular velocity, $\Omega_\mathrm{p}$ for spiral pattern speed, $\kappa$ for epicyclic frequency. For each pattern, a resonance of order $m$ is described by a curve $\Omega + \kappa/m$. A 1:1 resonance (hereafter CR) is defined by $\Omega_\mathrm{p}=\Omega$, i.e. material component (stars and gas) \textit{co-rotate} with density wave. A 2:1 resonance is called a Lindblad resonance, and we discern the inner one (ILR, $m=-2$) and the outer one (OLR, $m=2$). These are the main resonances, encircle the limits where the density wave can propagate \citep{1972MNRAS.157....1L}. Besides CR, ILR and OLR, we  mention the inner and outer 4:1 resonances, also referred to as 'ultraharmonic' resonances. We use the abbreviations IUHR and OUHR for them, where $m$ is equal $-4$ and $4$, respectively. Resonances of higher order values of $m$ are rarely used in studies of the dynamics of real galaxies and, thus, they are not considered in this analysis. We note that all resonance notations used here represent a radial position of that particular resonance.
\par
To date, at least five different cases are known where two resonances of inner (in) and outer (out) patterns overlap, possibly forming a non-linear coupling. All have been shown to occur in real galaxies, and often in numerical $N$-body simulations, and are listed below. First is $\mathrm{CR_{in} = ILR_{out}}$, predicted by \citet{masset1997} and found, for example, for IC342 in Fig.~8 in \citet{2009ApJ...702..277M}. The second is $\mathrm{CR_{in} = IUHR_{out}}$, demonstrated by \citet{2008ApJ...688..224M} for M51 and supported by \citet{2013ApJ...771...59M} findings. The third is $\mathrm{OLR_{in} = ILR_{out}}$, demonstrated for NGC4736 \citep{1995A&A...301..359M}. The fourth is $\mathrm{OLR_{in} = IUHR_{out}}$, as found to be the case in M101 \citep{2009ApJ...702..277M} and in NGC3433 \citep{2018ApJ...854..182B}. The fifth is  $\mathrm{OLR_{in} = CR_{out}}$, which was found in NGC3433 by \citet{2018ApJ...854..182B}. Curiously, the latter should also be the case for \citet{1989MNRAS.240..991S,2019MNRAS.489..116S} groove instability cycle, where each next mode is generated exactly as it's OLR is congruent with the previous CR. 
\par
The seminal work of \citet{1999A&A...348..737R} demonstrates the possibility of all of these cases except the forth, both for bar and spiral, and for spiral-spiral coupling. In \citet{2014ApJS..210....2F} authors study change of radial velocity sign in the kinematics of 104 galaxies using H${\alpha}$ data. They find the same repeating pattern in over 70\% of their sample, which is $\mathrm{CR_{in} = IUHR_{out}}$ and $\mathrm{OLR_{in} = CR_{out}}$ simultaneously. The coincidence has been found up to four times in one galaxy, and is repeated at least twice in many objects. Theoretically, their method is able to detect all the main resonances, but it is difficult to discriminate between them, and often there are several possible interpretations. 
\par
In principle, other forms of coupling are possible, at least formally. The following analysis will be constrained not only to the cases listed, which have already been shown to hold in real galaxies, but to all combinations of main resonances, presented in  Table~\ref{tab:main}. We naturally assume that when $\mathrm{CR_{in} < CR_{out}}$, then every resonance from the sequence ILR-IUHR-CR-OUHR-OLR of the inner pattern can be formally coupled with all those resonances of the outer pattern, which are placed to the left of it in this sequence. It is also important to note that coupling with bar is better studied, but we focus on the spiral-spiral case, and most or all of the above examples have been shown to be possible for spirals as well.

\subsection{Relative corotation radii positions}
\label{sec:crposition}

In this subsection, we will investigate the possible relative CR positions for consequent patterns, assuming resonance coupling. Let the rotation curve (RC) have the form $v(r) \propto r^{\alpha}$ and the equation for the epicyclic frequency, $\kappa$, to be: 

$$\kappa^2 = \frac{2\Omega}{r} \mathrm{\frac{d}{dr}} \left(r^2\Omega\right).$$

Then, it is trivial to find the position of any resonance $\mathrm{R_{m}}$, where $m$ values are listed for particular resonances in Sect.~\ref{sec:first}, relative to CR of the same pattern:

\begin{equation}
\frac{\mathrm{R_{m}}}{\mathrm{CR}} = \left(1 + \frac{\sqrt{2}}{m}\sqrt{1+\alpha}\right)^{(1-\alpha)^{-1}}.\label{eq:rcr}
\end{equation}

The Eq.~\ref{eq:rcr} can be found in \cite{1992ApJS...79...37E} and \cite{1995ApJ...445..591E}\footnote{Note: the authors accidentally missed the factor of 2 in the equation.}, but for the inverse sign of $m$. We apply this equation to all  observed coupling cases, assuming the case of flat RC: $v(r)=v_0=\mathrm{const}$, $\alpha=0$. Such a choice is motivated firstly  by the simplicity of Eq.~\ref{eq:rcr} in this case and, secondly, by the findings of \citet{1999A&A...348..737R}, who estimated that mode coupling seems to be the strongest when the halo contribution to the RC is large. Of course, real RCs are often flat in parts where spiral arms are hosted, but not in the central region. 
\par
Under the selected assumption, we get a simple expression:
\begin{equation}
\frac{\mathrm{R_{m}}}{\mathrm{CR}} = \left(1 + \frac{\sqrt{2}}{m}\right).\label{eq:rcr2}
\end{equation}
For each of the observed coupling cases, we solve a trivial set of equations. By defining $\upvarpi=\mathrm{CR_{out}}/\mathrm{CR_{in}}$ as a ratio of the inner CR to the outer CR ($\upvarpi > 1$) we get ten different values, which are presented in Table~\ref{tab:main}. For example, $\mathrm{OLR_{in} = ILR_{out}}$ congruence, shown to be the case in NGC4736, gives $\upvarpi=(1+\sqrt2/2)/(1-\sqrt2/2)\approx5.83$, and other cases are calculated in a similar way. We note that for flat RC  $\upvarpi^{-1}$ is equal to the ratio of pattern speeds $\mathrm{\Omega^{out}_{p}/\Omega^{in}_{p}}$.
Such simple considerations produce nevertheless several findings.

\begin{table}
        \centering
        \caption{All possible combinations and ratio $\upvarpi$ for main resonances.}
        \label{tab:main}
        \begin{tabular}{c|cccc} % four columns, alignment for each
                \hline
                   \diagbox{out}{in}   & IUHR & CR & OUHR & OLR \\
        % \noalign{\smallskip}
                \hline
  \noalign{\smallskip}
            ILR  & 2.21 & 3.41 & 4.62 & 5.83 \\
            IUHR  & - & 1.55 & 2.09 & 2.64 \\
            CR  & - & - & 1.35 & 1.71 \\
            OUHR  & - & - & - & 1.26 \\
                \hline
        \end{tabular}
\end{table}

\par
First, we demonstrate that in case of flat RC, regardless of whether it is bar or spirals, we only get a limited set of possibilities where the next pattern could be found.
Secondly, \citet{2014ApJS..210....2F} findings of simultaneous couplings coexisting naturally appear in the presented case, because $\upvarpi$ is close for them. Since $\Omega$ is subject to RC errors, and given that the radial extent of resonant orbits can be quite large, for all $\upvarpi=1.62\pm0.08$ or even for a larger range, we will find $\mathrm{OLR_{in} = CR_{out}}\; \textit{\&} \;  \mathrm{CR_{in} = IUHR_{out}}$ within the errors.
Third, it is easy to see that a similar situation holds for two other pairs of coupling: for $\upvarpi=1.31\pm0.04$ we simultaneously observe $\mathrm{OUHR_{in} = CR_{out}} \; \textit{\&} \;\mathrm{OLR_{in} = OUHR_{out}}$, and for 
$\upvarpi=2.15\pm0.06$ we will find $\mathrm{OUHR_{in} = IUHR_{out}}$ \; \textit{\&} \; $\mathrm{IUHR_{in} = ILR_{out}}$. In both cases, the uncertainty is smaller than in the case of \citet{2014ApJS..210....2F}, namely, the apparent overlaps are spatially less extensive and, to the best of our knowledge, this result has not been noticed before. 
\par
Finally, we see that for all the remaining cases there is almost no room for a second pattern in the disc for flat RC and spiral-spiral coupling. In these cases, if the $\mathrm{CR_{in}}$ is further than $0.3r_{25}$ from the center, which is observed in a significant fraction of galaxies \citep{1995ApJ...445..591E}, then $\mathrm{CR_{out}}$ is far beyond the optical radius. Theoretically, this could be possible if the outer spiral extends significantly beyond the disc (e.g. \citealp{2024MNRAS.52710615M}) or if it lies entirely inside the CR, meaning that it rotates faster than the disc. Both of these possibilities are hard to imagine being frequently occurred, making the discussed coupling unlikely. 

\subsection{Winding time}
\label{sec:wind}

The problem of winding of spiral arms have been recognised for some time. Formulated by \citet{oort} as the `winding dilemma', it states that if spiral arms were material features, they would wind more and more and tightly, until they finally disappear, on timescales of $\sim100$~Myr. Advances in density wave theory, as detailed in \citet{1966PNAS...55..229L} and other works (e.g. see \citealp{2016ARA&A..54..667S} and \citealp{2005AAS...207.3101M}) allow this puzzle to be solved. However, in the presence of two or more separate spiral patterns, as in the resonance coupling cases considered, the same dilemma reappears: the inner part inevitably rotates faster than the outer, and the tips of the spirals should diverge from each other, forming bifurcations or gaps in the arms.
\par
Following \citet{2006MNRAS.366L..17M}, we define the winding time $\tau_{wind}$ as the time needed for the patterns to diverge by one full rotation, namely, $\tau_{wind} = 2\pi/\large(\Omega_{\mathrm{max}} - \Omega_{\mathrm{min}}\large)$. Under the given assumptions, it should then be transformed into:

$$\tau_{wind} = 2\pi\left(\frac{v_0}{\mathrm{CR_{in}}} - \frac{v_0}{\mathrm{CR_{out}}}\right)^{-1} = \frac{2\pi}{\Omega_{\mathrm{in}}} \times \left(1 - \frac{{\mathrm{CR_{in}}}}{{\mathrm{CR_{out}}}}\right)^{-1} = \frac{\tau_{GY}}{1 - \upvarpi^{-1}},$$ 

where $\tau_{GY}$ is a time of one disc rotation (or 'galactic year') in a place of the inner pattern CR. Therefore, for $\upvarpi$ values from Table~\ref{tab:main} we get values of $\tau_{wind}$ between $1.2\tau_{GY}$ and $4.8\tau_{GY}$. The obtained results show that for most of the presented cases of resonance coupling, the global spiral pattern should wind up in less than three rotations of the disc, if all individual patterns are stationary, and often even twice faster. In \citet{masset1997} authors had already noticed this for bars, that for different pattern speeds the tips of the bar and beginnings of the spirals should be found at random relative azimuthal positions. At the same time, simulations of \citet{minchev2012} and others demonstrate a preserved and consistent pattern regardless of the coupling in the system. It was explained in \citet{minchev2012} that each time the bar encounters the spiral, the inner wave is regenerated, accelerating to catch up with the spiral. It is likely that a similar behavior may work for the spiral-spiral case or perhaps spirals are transient in nature rather than long-lived, as \citet{2006MNRAS.366L..17M} concluded, and follow some form of recurrence \citep{1989MNRAS.240..991S}.

\section{Comparison with observations}

We go on to demonstrate the feasibility of obtained results using observations for six nearby galaxies, listed in Table~\ref{tab:galparams}. These are well-known and well-studied objects of late Hubble types and intermediate inclinations. We note that the conducted analysis is not dependent on the exact inclination and PA adopted. All galaxies demonstrate a multi-arm morphology, often with two main arms of greater contrast, as seen in images from the DESI Legacy Imaging Surveys \citep{2019AJ....157..168D} in Figure~\ref{fig:thumbnail}. According to the literature, only NGC3344 and NGC5371 contain a noticeable bar.
\par
Due to the importance of CR estimations for processes related to bars and spirals in galaxies, there are dozens of various methods for estimating it. These include a direct Tremaine-Weinberg method \citep{1984ApJ...282L...5T,2009ApJ...704.1657F,2021AJ....161..185W}, with modifications (generalisation to multiple pattern speeds in \citealp{2009ApJ...702..277M}), a measurement of the shift between potential and density \citep{2009ApJS..182..559B}, various angular offset-based methods \citep{1990ApJ...349..497C,2008AJ....136.2872T,2020MNRAS.496.1610A,2021MNRAS.508..912S}, detections of  morphological peculiarities \citep{1975ApJ...196..381R,1992ApJS...79...37E,1995ApJ...445..591E}, $N$-body modelling \citep{2003ApJ...586..143K}, abundance gradients \citep{1992MNRAS.259..121V,2013MNRAS.428..625S},  and radial velocity sign changes \citep{2011ApJ...741L..14F,2014ApJS..210....2F}. These are not all approaches to CR detection available, with new ones  reported regularly \citep{2024MNRAS.527L..66M}, but the absolute majority of works use a selection of them. 
\par
The values of CR, estimated in different works, tend to be inconsistent with each other \citep{kost}. We present reliable CR estimates for the galaxies under consideration in Fig.~\ref{fig:crs} and in Table~\ref{tab:crref}. For each galaxy we have at least three different CR measurements, obtained by at least three different methods, using all the aforementioned works. For NGC2403, NGC3893, and NGC5676 we found data to be consistent with three different CRs, and for three other galaxies we find two. We note that each CR, except the first one for NGC2403, has been confirmed by at least two different measurements. This fact, together with the visible agreement of individual values obtained using both ISM and stellar data sources with different photometric bands, significantly increases the certainty and reliability of the CR estimate.
\par
Individual RCs and their approximations are shown in Fig.~\ref{fig:rotvels}. For each galaxy, we find HI observations in order to track cold circular velocity, in some cases backed with additional observations. 
% We fit data points with function $v_0\times r^2/(\mathrm{const} + r^2)$ scaled to actual plateau velocity $v_0$. 
Indeed, it is easy to see from Fig.~\ref{fig:rotvels} that all RCs are flat over the range of radii considered to be important in this work, but with two sidenotes. Firstly, we note that the first CR in NGC3893 is located before $v(r)$ reaches the plateau velocity, $v_0$. Secondly, in NGC5371, data points do not agree perfectly with each other resulting in what is probably a larger $v_0$ value than in the fit, but all this does not greatly affect  the colored areas shown in Fig.~\ref{fig:rotvels}. We note that all CRs are located within the canonical optical radius $r_{25}$, found in \citet{2014A&A...570A..13M}.

We present the frequency curves $\Omega(r)$ and their associated resonances in Fig.~\ref{fig:omegas}. It is easy to see that all 6 galaxies demonstrate the apparent pattern coupling, visible in this Figure. In all cases (except NGC2403, where it is also presented, but not for subsequent patterns), we see the same repetitive situation: OLR of the inner pattern coincides with CR of the next, and CR of inner lies within the radius of the outer IUHR. For one galaxy, NGC5675, we even see it repeated twice for three separate spiral patterns. We note that the ratio $\upvarpi$ between corotation radii, shown in Fig.~\ref{fig:omegas}, is in all cases in a good agreement within uncertainties with predictions listed in Table~\ref{tab:main} and the predictions from Sect.~\ref{sec:crposition}, along with the same occurrence of according couplings. 
\par
Several insights come from Fig.~\ref{fig:omegas}. First, the case of special interest is the galaxy NGC2403 and the first resonance coupling in NGC3893. Here, we see the case where $\mathrm{OLR_{in}=OUHR_{out}}$ and $\mathrm{OUHR_{in}=CR_{out}}$, neither of which has been previously reported as far as we aware. Their synchronous appearance is not accidental and also shows agreement with Sect.~\ref{sec:crposition}, which gives exactly the predicted $\upvarpi$ for NGC2403, where it is repeated twice. This is not the case for NGC3893, because the first CR is not located within the RC plateau. Secondly, we notice that for galaxies with more than two patterns, the central CR is coupled with both of them. This is very similar to the case of NGC3433 presented in \citet{2018ApJ...854..182B}, where the central pattern is also supported from both sides. Finally, if we also mark the bar size, which is 25.5~arcsec for NGC3344 and 19.3~arcsec for NGC5371 \citep{2015A&A...582A..86H}, and taking into account that the bar ends near its CR, we find that the bar is coupled with ILR of spiral pattern for both galaxies.

\section{Discussion, additional notes, and conclusion}

For several of the galaxies included in this study, we are not the first to report resonance coupling. As already mentioned, \citet{2011ApJ...741L..14F,2014ApJS..210....2F} studied changes of radial velocity sign and found them in many cases, including the  presented NGC3344, NGC3893, and NGC5676 (UGC5840, UGC6778 and UGC9366 in their works). Along with other sources, we also used their CRs measurements in these galaxies too, as indicated in Table~\ref{tab:crref}. Therefore, the results for these galaxies may seem trivial at first sight. However, firstly, we used other sources and methods to confirm the measurements and, secondly, interpretation plays a crucial role here. For example, \citet{2014ApJS..210....2F} for NGC3344 interpret last CR as OLR of the second pattern, without mentioning the coupling. For NGC3893, they did not notice the coupling at $63\pm3$~arcsec and interpreted the last point as a separate pattern, when it should be OUHR. Finally, in \citet{2014ApJS..210....2F} authors did not detect the important last CR we found in NGC5676 and their $47.2\pm2.6$~arcsec point is likely not another CR; instead it is an OUHR, as Fig.~\ref{fig:omegas} clearly shows. In summary, we show that even in these galaxies, the reported coupling has not been fully recognised.
\par

\citet{2011ApJ...735..101F} studied angular offsets between different indicators using the cross-correlation method to find that they are not consistent with a stable density wave, including NGC628 and NGC2403, which we analyse here. However, \citet{2014ApJS..210....2F} found that measured offsets are actually broadly consistent with density wave, assuming the existence of multiple patterns. Indeed, we see the expected zero angular offsets in radii close to CRs obtained in NGC628 and for the first CR in NGC2403, where the \citet{2011ApJ...735..101F} analysis is not sufficiently extended. In any case, the application of the cross-correlation method to azimuthal profiles is difficult for galaxies with complex morphologies.
\par
We also want to mention some supporting evidence that indicates the estimated CRs are correct. For example, we can see that the angular offsets measured by \citet{2008AJ....136.2872T} for NGC2403 could easily be fitted by two curves, or that the metallicity values in \citet{2013ApJ...775..128B} for the same galaxy show a break at the location of the first CR. For NGC5371, we find that the residuals of the velocity field for the HI data in \citet{1987PhDT.......209B} are consistent with Fig.~\ref{fig:crs}, when we apply the \citet{2011ApJ...741L..14F} method.
\par
Coupling could also leave an imprint on the galactic morphology. For all galaxies, we see a complex morphology, that there are several long extended arms and many that diverge from them, which look shorter and weaker (Figure~\ref{fig:thumbnail}). \citet{2014ApJS..210....2F} called this ``organised pseudo-flocculence.'' Similarly to our case, we are able to see clear bifurcations for NGC3433 in \citet{2018ApJ...854..182B} in their Fig.~1. The reason is probably the winding, as demonstrated in Sect.~\ref{sec:wind}, or it is due to the existence of beat waves \citep{masset1997,1992ApJS...79...37E}. The simulations from \citet{2011MNRAS.417..762Q} and \citet{minchev2012} clearly show gaps and discontinuities in the main spiral arms, arguing that they indicate changes in the dominant pattern, namely, the transition between the inner and outer structures. Finally, the Landau damping mechanism could also be the reason, as \citet{1996ApJ...472..532M} suggested. 
\par
Another feature of a galaxy's appearance that is relevant to his discussion consists of the breaks in their discs. \citet{minchev2012} and \citet{2012MNRAS.426.2089R} used numerical models to demonstrate that resonance overlap leads to the eventual formation of a break at the position of CR. This is exactly what could be found, for example, for NGC3893. In the case, we see a remarkable coincidence between the estimated CRs (see Fig.~\ref{fig:crs})  of individual patterns and the change on the exponential scale, presented on the azimuthally averaged profile from \citet{2015ApJS..219....4S}. Another  interesting piece of evidence that may support this theory comes from \citet{2013ApJ...771...59M}, who found that many galaxies have disc breaks at about $3.5R_{\mathrm{bar}}$, where $R_{\mathrm{bar}}$ is a half-size of a bar major axis. These authors have made similar arguments about the flat RC and interpretation that the break occurs at the spiral's OLR due to coupling. However, we want to note that it can alternatively be explained by the fact that $R_{\mathrm{bar}}\approx R_{\mathrm{CR}}$ and, according to Table~\ref{tab:main}, we can observe that breaks form exactly at the spiral's CR, which is in better agreement with \citet{minchev2012} and \citet{2012MNRAS.426.2089R} predictions. Certainly, the question under consideration is more difficult than that, because we can find CRs exactly where spirals form ``bump''  on top of the disc \citep{2011MNRAS.414..538K,2024MNRAS.528.1276M} or the breaks themselves could be the reason for several patterns and apparent coupling  \citep{2024MNRAS.529.4879F}. This topic needs a separate and careful investigation.

\par
The reliable estimations of CR or equivalent $\Omega_\mathrm{p}$ presented in this work are important in and of themselves for a number of reasons, including the crucial role of CR in the swing amplification mechanism \citep{1981seng.proc..111T}, its connection with orbital \citep{2013MNRAS.436.1201C} and gravitational \citep{2021MNRAS.506...84I} stability, chemical evolution \citep{1992MNRAS.259..121V} in the disc, and other issues, such as local star formation \citep{2022ApJ...941L..27W}. In addition, we would like to emphasise that resonance coupling may be of great importance not only for angular momentum transfer \citep{2002MNRAS.336..785S,masset1997,minchev2012}, but also for disc heating \citep{2006MNRAS.368..623M} and for magnetic field generation on the galactic scale, as suggested by  \citet{2014MNRAS.437..562C}. Another important application of the results obtained relates to our own Galaxy. \citet{2003NewA....8...39S} summarised the $\Omega_{\mathrm{p}}$ measurements for MW in their Table~3, where the measurements are clearly concentrated in two sets with average about 22--26~km/s/kpc and 14~km/s/kpc (see also \citealp{2022AstL...48..568B,2021MNRAS.506..523V,2007NewA...12..410N}). Given that RC of Galaxy to a first approximation is flat here \citep{2017A&A...601L...5R}, we obtained $\upvarpi\approx1.7$, namely, MW show signs of the ongoing coupling between patterns as found in \citet{2014ApJS..210....2F}.
\par
The nature of spiral arms remains elusive en masse. In this preliminary study, our conclusions are as follows.

      (i) Assuming a flat rotation (RC) curve and spiral-spiral resonance coupling, we find the ratio $\upvarpi=\mathrm{CR_{out}}/\mathrm{CR_{in}}$ of corotation radii (CRs) of two consequent patterns for all main resonances overlapping, presented in Table~\ref{tab:main}. In three cases, we get $\upvarpi$ values greater than 3, so there is barely enough space for two or more spiral patterns in the disc. We thus predict that examples of such galaxies are expected to be rare, if they even exist. 
      \par
      (ii) We demonstrate that simultaneous coupling $\mathrm{OLR_{in}=CR_{out}}\; \textit{\&} \;\mathrm{CR_{in}=IUHR_{out}}$, observed in many galaxies  by \citet{2014ApJS..210....2F}, appears very naturally in our formulation due to the very close $\upvarpi$ ratio in both cases. This is also true for two other pairs, mentioned in Sect.~\ref{sec:crposition}.
      \par
      (iii) For six galaxies with flat RC, we estimate several corotation radii, using measurements from other works (Table~\ref{tab:crref}). For each galaxy, we used at least three different independent methods to increase the level of confidence. We predict that new accurate measurements will fit with the presented CRs and found supporting evidence. 
      \par
      (iv) For these galaxies, we demonstrate that the estimated observational resonances are visually coupled (Fig.~\ref{fig:omegas}) and agree with expectations, as well as the $\upvarpi$ ratio. This is not the first time, when coupling has been demonstrated in real galaxies, but it is now well motivated. The substantial arguments found for resonance coupling provide strong observational evidence for the existence of several individual spiral patterns simultaneously in one galaxy.
      \par
      (v) We demonstrate that resonance coupling inevitably means that spirals are expected to 'wind up' in several rotations. We estimate the wind-up time $\tau_{wind}$ for each case in terms of rotations. 
      \par
      (vi) We find a new resonance coupling variant in NGC2403 (twice) and NGC3893, specifically $\mathrm{OLR_{in}=OUHR_{out}}\; \textit{\&} \;\mathrm{OUHR_{in}=CR_{out}}$. We believe that both separate and simultaneous cases have not been noticed before and, similarly to the case of \citet{2014ApJS..210....2F}, this coincidence ought to be expected from the proximity of $\upvarpi$.

In a future work, we will continue to explore the dynamics and nature of spiral arms by extending this analysis to new objects. Thus, we plan to focus on the galaxy M109 (NGC3992), which has a flat RC and obvious coupling, but with only one measurement for each CR. Another is M101 (NGC5457), where we see the same case of coupling repeated three times according to the measured $\upvarpi$, but where the HI RC is limited and contains only the first CR. In total, we have at least ten less reliable candidate galaxies with probable observed resonance coupling. We will also generalise the obtained result for non-flat RC and take a closer examination of the interplay between morphological features (number of arms, disc breaks) and properties of individual spiral patterns.

\bibliographystyle{aa}

\begin{appendix}
\section{Observational data}

\begin{figure}\includegraphics[width=0.95\columnwidth]{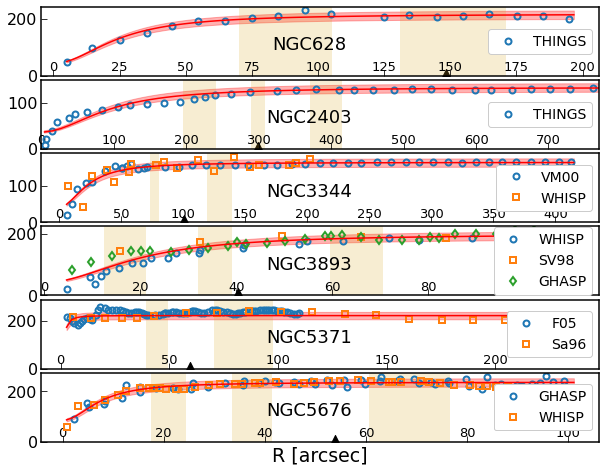}
    \caption{Rotation curves (markers) and their fits (solid line) with 7\% relative error. Triangle symbol marks $r_{25}/2$ position. References: F05 (H{$\alpha$}): \citet{2005A&A...430...67F}; THINGS (HI): \citet{2008AJ....136.2648D}; GHASP (H{$\alpha$}): \citet{2005MNRAS.362..127G}; SV98 (HI): \citet{1998ApJ...503...97S};  VM00 (optical): \citet{2000A&A...356..827V}; WHISP (HI): \citet{2001ASPC..240..451V}; and Sa96 (HI): \citet{1996ApJ...473..117S}.}
    \label{fig:rotvels}
\end{figure}

Here in Table~\ref{tab:crref} we list corotation measurements compiled from the literature for six galaxies and references for them, and in Table~\ref{tab:galparams} present main properties of used galaxies. In Figure~\ref{fig:rotvels} we show rotation curves, compiled from the literature sources.

\begin{table}
        \centering
        \caption{Parameters of galaxies. References are: (i) \citet{2018MNRAS.476.3591M}; (ii) \citet{2023MNRAS.520..147L}; (iii) \citet{2009ApJ...702..277M}; (iv) \citet{2014ApJS..210....2F}; (v) \citet{2009ApJ...704.1657F}. }
        \label{tab:galparams}
        \begin{tabular}{lccclc} % four columns, alignment for each
                \hline
                Name & incl. & PA & Dist. & Ref. & $r_{25}$\\
                - & deg & deg & Mpc & - & arcsec\\
            \noalign{\smallskip}
                \hline
  \noalign{\smallskip}
                NGC\,628 & $7$ & $20$ & 8.6 & (i) & 296.6\\
                NGC\,2403 & $63$ & $125$ & 3.2 & (ii) & 598.6\\
                NGC\,3344 & $25$ & $155$ & 6.9 & (iii) & 201.0\\
                NGC\,3893 & $49$ & $343$ & 15.5 & (iv) & 80.7\\
        NGC\,5371 & $48$ & $0\pm12$ & 37.8 & (v) & 119.4\\
        NGC\,5676 & $62$ & $225$ & 37.7 & (iv) & 107.7\\
                \hline
        \end{tabular}
\end{table}

\begin{table}
        \centering
        \caption{CR measurements and reference works: [1]  \citet{1995ApJ...445..591E}; [2]  \citet{2009ApJ...702..277M}; [3]  \citet{2011ApJ...741L..14F,2014ApJS..210....2F}; [4]  \citet{2008AJ....136.2872T}; [5]  \citet{2013MNRAS.428..625S}; [6]  \citet{2009ApJ...704.1657F}; [7] \citet{2009ApJ...694..512M}; [8]  \citet{1975ApJ...196..381R}; [9]  \citet{2003ApJ...586..143K}; [10]  \citet{2021AJ....161..185W}; [11]  \citet{2020MNRAS.496.1610A}; [12]  \citet{2009ApJS..182..559B}; [13] \citet{1992MNRAS.259..121V}; [14]  \citet{1990ApJ...349..497C}; [15]  \citet{2021MNRAS.508..912S}; [16]  \citet{1992ApJS...79...37E}; [17] \citet{2024MNRAS.527L..66M}.}
        \label{tab:crref}
        \begin{tabular}{lll|lll} % four columns, alignment for each
                \hline
                NGC & CR, arcsec & Ref. & NGC & CR, arcsec & Ref.\\
  \noalign{\smallskip}
                \hline
            \noalign{\smallskip}
628  &  $80.2\pm25.3$  & [10] & --///--  &  $79.7\pm1.6$  & [3]\\
--///--  &  $85.9\pm28.6$  & [15] & --///--  &  $119.6\pm12.9$  & [2]\\
--///--  &  $85.9\pm28.6$  & [15] & --///--  &  $120.6\pm1.6$  & [3]\\
--///--  &  $88.8\pm4.2$  & [1] & --///--  &  $127.3\pm2.3$  & [3]\\
--///--  &  $91.3\pm20.6$  & [10] & --///--  &  $142.0\pm8.0$  & [17]\\\cline{4-6}
--///--  &  $94.5\pm42.8$  & [10] & 3893  &  $13.0\pm6.2$  & [3]\\
--///--  &  $137.0\pm20.9$  & [11] & --///--  &  $20.8\pm0.0$  & [12]\\
--///--  &  $138.5\pm20.9$  & [11] & --///--  &  $34.9\pm4.2$  & [3]\\
--///--  &  $141.0\pm0.0$  & [14] & --///--  &  $36.3\pm2.7$  & [1]\\
--///--  &  $141.3\pm0.0$  & [16] & --///--  &  $61.0\pm0.0$  & [12]\\
--///--  &  $145.2\pm40.4$  & [4] & --///--  &  $63.0\pm3.0$  & [3]\\
--///--  &  $147.8\pm55.4$  & [5] & --///--  &  $66.7\pm6.1$  & [9]\\\cline{4-6}
--///--  &  $177.6\pm28.6$  & [15] & 5371  &  $44.3\pm6.1$  & [7]\\
--///--  &  $180.5\pm28.6$  & [15] & --///--  &  $44.7\pm2.1$  & [1]\\\cline{1-3}
2403  &  $218.4\pm23.4$  & [4] & --///--  &  $66.6\pm13.3$  & [7]\\
--///--  &  $294.0\pm0.0$  & [13] & --///--  &  $93.0_{-16.0}^{+0.0}$  & [6]\\
--///--  &  $304.6\pm0.0$  & [8] & --///--  &  $92.0\pm10.0$  & [17]\\
--///--  &  $390.0_{-32.0}^{+0.0}$  & [7] & 5676  &  $18.4\pm1.9$  & [3]\\\cline{4-6}
--///--  &  $392.3\pm35.7$  & [5] & --///--  &  $21.3\pm1.6$  & [3]\\\cline{1-3}
3344  &  $74.4\pm1.9$  & [3] & --///--  &  $23.2\pm0.0$  & [12]\\
--///--  &  $78.0\pm5.4$  & [1] & --///--  &  $36.9\pm5.6$  & [3]\\
--///--  &  $79.7\pm1.6$  & [3] & --///--  &  $38.2\pm0.0$  & [12]\\
\hline
        \end{tabular}
\end{table}

\begin{figure}\includegraphics[width=0.95\columnwidth]{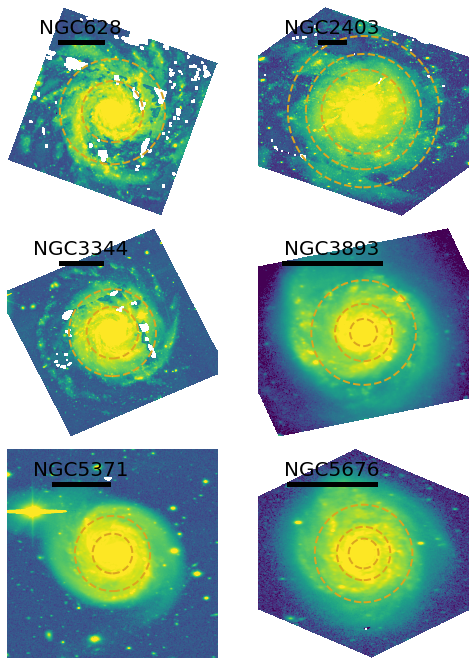}
    \caption{Thumbnail images of galaxies from DESI Legacy Imaging Surveys \citep{2019AJ....157..168D} in the $g$~band. Images are in logarithmic scale  in arbitrary units, circles show the average positions of each CR accordingly, and the scalebar length in each frame is equal to 2~arcmin.}
    \label{fig:thumbnail}
\end{figure}

\end{appendix}
\end{document}